\documentstyle[preprint,aps]{revtex}

\begin{document}
\title{Performing Quantum Measurement in Suitably Entangled States Originates the
Quantum Computation Speed Up}
\author{Giuseppe Castagnoli\thanks{%
Elsag Bailey and Universit\`{a} di Genova, 1654 Genova, Italy}, Dalida Monti%
\thanks{%
Elsag Bailey and Universit\`{a} di Genova, 1654 Genova, Italy}, and
Alexander Sergienko\thanks{%
Dept. of Electrical and Computer Engineering, Boston University, Boston, MA
02215, USA}}
\date{\today}
\maketitle

\begin{abstract}
We provide a justification of the quantum speed-up based on the
complementary roles played by the reversible preparation of an entangled
state before measurement and by the final measurement action.
\end{abstract}

\section{Introduction}

\noindent Why quantum computation can be more efficient than its classical
counterpart is an open problem attracting increasing attention\cite{EKJ98}, 
\cite{KIT}. The reason is naturally sought in the special features of
quantum mechanics exploited in quantum computation, like state
superposition, entanglement and quantum interference. Quantum measurement,
instead, is generally considered necessary only to ``read'' the computation
output. In the justification we shall provide, measurement does more than
``reading'' an output, it contributes in creating that output in a
computationally efficient way.

We will show that the logical constraint that there is a single measurement
outcome, acquires a striking function in existing quantum algorithms. It
becomes a set of logical-mathematical constraints representing the problem
to be solved, or the hard part thereof, whereas the measurement outcome, by
satisfying these constraints, yields the solution.

In all these algorithms, the state before measurement is entangled with
respect to a couple of observables\footnote{%
As we will see, also in Deutsch's and Grover's algorithms, provided that
both the problem and the solution algorithm are represented in a physical
way.}. It is a basic axiom of quantum measurement theory that the time
required to measure an observable is independent of this possible
entanglement: entanglement is interaction-free.\ The computational
complexity of satisfying the above logical-mathematical constraints
originates from entanglement and is transparent to measurement time.

On the basis of these arguments, we will justify the speed-up in all known
quantum algorithms.

\section{Overview}

\noindent For unity of exposition, we shall provide an overview of our
justification of the speed-up based on a simplified version of Simon's
algorithm. All details are deferred to the subsequent Sections.

The problem is as follows. Given $B=\left\{ 0,1\right\} $, we consider a
function $f\left( x\right) $ from $B^{n}$ to $B^{n}$. The argument $x$
ranges over $0,1,$ $...,$ $N-1$, where $N=2^{n}$; $n$ is said to be the size
of the problem.

We assume that $f\left( x\right) $ has the following properties:

\begin{itemize}
\item  it is a 2-to-1 function, namely for any $x\in B^{n}$ there is one and
only one second argument $x^{^{\prime }}\in B^{n}$ such that $x\neq
x^{^{\prime }}$ and $f\left( x\right) =f\left( x^{^{\prime }}\right) $;

\item  such $x$ and $x^{^{\prime }}$ are evenly spaced by a constant value $%
r $, namely: $\ \left| x-x^{^{\prime }}\right| =r$;

\item  given a value $x$ of the argument, computing the corresponding value
of $f\left( x\right) $ requires a time polynomial in $n$ [poly$\left(
n\right) $]; whereas, given a value $f$ of the function, finding an $x$ such
that $f\left( x\right) =f$, requires a time exponential in $n$ [exp$\left(
n\right) $]; the function is ``hard to reverse''.
\end{itemize}

Besides knowing the above properties, we can use a quantum computer that,
given any input $x$, produces the output $f\left( x\right) $ in poly$\left(
n\right) $ time. The problem is to find $r$ in an efficient way, which turns
out to be in poly$\left( n\right) $ rather than exp$\left( n\right) $ time.

The computer operates on two registers $a$ and $v$, each of $n$ qubits; $a$
contains the argument $x$ and $v$ -- initially set at zero -- will contain
the result of computing $f\left( x\right) $. We denote by ${\cal H}%
_{av}\equiv span\left\{ \left| x\right\rangle _{a},\left| y\right\rangle
_{v}\right\} $, with $\left( x,y\right) $ running over $B^{n}\times B^{n}$,
the Hilbert space of the two registers.

By using the quantum computer and standard operations like the Hadamard
transform (see IV\ for details), we obtain in poly$\left( n\right) $ time,
at time $t_{2}$, the following state of the two registers (indexes are as in
IV):

\begin{equation}
\left| \varphi ,t_{2}\right\rangle _{av}=\frac{1}{\sqrt{N}}\sum_{x}\left|
x\right\rangle _{a}\left| f\left( x\right) \right\rangle _{v},
\end{equation}

\noindent with $x$ running over $0,1,$ $...,$ $N-1$.

We designate by $\left[ a\right] $ (an observable) the number stored in
register $a$. Similarly $\left[ v\right] $ is the number stored in $v$. We
measure $\left[ v\right] $ in state (1)\footnote{%
This intermediate measurement can be skipped, but we will see that it is
mathematically equivalent either performing or skipping it.}. Given the
character of $f\left( x\right) $, measurement outcome has the form:

\begin{equation}
\left| \varphi ,t_{3}\right\rangle _{av}=\frac{1}{\sqrt{2}}\left( \left| 
\overline{x}\right\rangle _{a}+\left| \overline{x}+r\right\rangle
_{a}\right) \left| \overline{f}\right\rangle _{v},
\end{equation}

\noindent where $\stackrel{\_}{f}$ is the value of the measured observable,
\noindent and $f\left( \overline{x}\right) =f\left( \overline{x}+r\right) =%
\overline{f}.$

We will see that, under a reasonable criterion, the quantum speed-up has
already been achieved by reaching state (2) -- see also Section IV.

Since the speed-up is referred to an efficient classical computation that
yields the same result, the quantum character of state (2) constitutes a
difficulty. This difficulty can be avoided by resorting to the notion of the
computational cost of classically producing the {\em description} of state
(2). This criterion yields a more universal way of comparing quantum and
classical efficiency, and coincides with the usual one when the quantum
algorithm has produced the ``classical reading''. It will be instrumental in
achieving an a-posteriori self-evident result.

Thus, we should assess the cost of classically producing description (2). Of
course, we must think that $\overline{x}$, $\overline{x}+r,$ and $\overline{f%
}$ are appropriate numerical values. Finding them requires solving the
following system of numerical algebraic equations:

\begin{eqnarray}
f\left( x_{1}\right) &=&f\left( x_{2}\right) , \\
x_{1} &\neq &x_{2}.  \nonumber
\end{eqnarray}

\begin{center}
Fig. 1
\end{center}

It is convenient to resort to the network representation of equations (3) --
fig. 1. The gate $c\left( x_{1},x_{2}\right) $ imposes that, if $x_{1}\neq
x_{2}$, then the output is 1, if $x_{1}=x_{2}$, then the output is 0, and
vice-versa. To impose $x_{1}\neq x_{2}$, the output must be set at 1. Note
that the network represents a system of algebraic equations: time is not
involved and gates are just logical constraints.

Each of the two gates $f\left( x\right) $ imposes that, if the input is $x$,
then the output is $f\left( x\right) $ or, conversely, if the output is $f$,
then the input is an $x$ such that $f\left( x\right) =f.$

This network is hard to satisfy by classical means. Because of the looped
network topology, finding a valuation of $x_{1},x_{2}$ and $f$ satisfying
the network requires reversing $f\left( x\right) $ at least once, which
takes, by assumption, exp$\left( n\right) $ time.

Instead, the time to produce state (2) with Simon's algorithm, is the sum of
the poly$\left( n\right) $ time required to produce state (1), and the time
required to measure the observable $\left[ v\right] $ in state (1). This
latter is independent of the entanglement between registers $v$ and $a$ and
is simply linear in the number of qubits of register $v$, namely in $n$. The
overall time is poly$\left( n\right) $. Under the above criterion, the
speed-up has already been achieved.

We shall provide two ways of seeing the active role played by the action of
measuring $\left[ v\right] $.

In Section III, we will show that measuring $\left[ v\right] $ introduces
and satisfies, in linear$\left( n\right) $ time, a system of algebraic
equations in Hilbert space (the one-outcome constraint and consequent ones)
equivalent, under the above criterion, to the system of algebraic equations
(3).

Here, this active role will be discussed at a conceptual level. The previous
criterion needs to be extended. The computational cost of producing a
quantum state starting from another quantum state will be benchmarked with
the cost of classically producing the description of the former starting
from the description of the latter.

We shall instrumentally use the following way of thinking (opposite to our
view):

\begin{quote}
{\footnotesize quantum computation can produce a number of parallel outputs
exponential in register size, at the cost of producing one output, but this
``exponential wealth'' is easily spoiled by the fact that quantum
measurement reads only one output.}
\end{quote}

Let us examine the cost of classically deriving description (2) from
description (1). The latter can be visualized as the print-out of the sum of
2$^{n}$ tensor products. Loosely speaking, two values of $x$ such that $%
f\left( x_{1}\right) =f\left( x_{2}\right) $, must be exp$\left( n\right) $
spaced. Otherwise such a pair of values could be found in poly$\left(
n\right) $ time by classical ``trial and error''.

The point is that the print-out would create a Babel Library\footnote{%
From the story ``The Library of Babel'' by J.L. Borges.} effect. Even for a
small $n$, it would fill the entire known universe with, say, $...$ $\left|
x_{1}\right\rangle _{a}\left| f\left( x_{1}\right) \right\rangle _{v}$ $...$
here, and $...$ $\left| x_{2}\right\rangle _{a}\left| f\left( x_{2}\right)
\right\rangle _{v}$ $...$ [such that $f\left( x_{1}\right) =f\left(
x_{2}\right) $] in Alpha Centauri. Finding such a pair of print-outs would
still require exp$\left( n\right) $ time. The capability of directly
accessing that ``exponential wealth'' would be vanified by its ``exponential
dilution''.

Quantum measurement, instead, {\em distills} the desired pair of arguments
in a time \ linear in $n$. In fact, it does more than randomly selecting one
measurement outcome; by selecting {\em one} outcome, it performs a logical
operation (selecting the two values of $x$ associated with the value of that
outcome) crucial for solving the problem.

The active role played by quantum measurement, complementary to the
production of the parallel computation outputs, is self-evident. In Section
III, this role will be pinpointed in a rigorous way.

\section{Quantum algebraic computation}

\noindent It is easy to show that quantum measurement introduces and
satisfies a system of algebraic equations equivalent to (3). By going
through elementary notions, we will highlight the pattern of a new form of
computation.

We shall first apply von Neumann's model to the quantum measurement of $%
\left[ v\right] $ in state (1). This model is two steps. The first is a
unitary evolution $U$, leading from the state before measurement to a
``provisional description'' of the state after measurement:

\begin{equation}
\left| \psi ,t_{2}\right\rangle _{avp}=\left| \varphi ,t_{2}\right\rangle
_{av}\left| 0\right\rangle _{p}\stackrel{U}{\longrightarrow }
\end{equation}

\begin{equation}
\left| \psi ,t_{3}\right\rangle _{avp}=\frac{1}{\sqrt{N}}%
\mathop{\textstyle\sum}%
_{i}\left( \left| x_{i}\right\rangle _{a}+\left| x_{i}+r\right\rangle
_{a}\right) \left| f_{i}\right\rangle _{v}\left| f_{i}\right\rangle _{p},
\end{equation}

\noindent where $f_{i}=f\left( x_{i}\right) =f\left( x_{i}+r\right) $. Here $%
p$ denotes a third register of $n$ qubits used to represent the state of the
``classical pointer'' in Hilbert space. This is sharp in state (4), before
measurement interaction. In the state after measurement (5), $f_{i}$ runs
over all the values of $f\left( x\right) .$ As stated before, the elapsed
time $t_{3}-t_{2}$ is linear in $n$ (the number of qubits in register $v$).
As well known, description (5) represents the appropriate entanglement
between measured observable and classical pointer, but it must be reconciled
with the empirical evidence that the pointer is in a sharp state.

The second step of von Neumann's model amounts to be a {\em reinterpretation}
of description (5). \ The tensor products appearing in (5) become mutually
exclusive measurement outcomes (still at the same time $t_{3}$) of
probability distribution the square modules of the respective probability
amplitudes, as well known\footnote{%
It is the same in decoherence theory, where the elements of a mixture become
mutually exclusive measurement outcomes.}. This yields a measurement outcome
of the form:

\[
\left| \varphi ,t_{3}\right\rangle _{av}\left| \overline{f}\right\rangle
_{p}=\frac{1}{\sqrt{2}}\left( \left| \overline{x}\right\rangle _{a}+\left| 
\overline{x}+r\right\rangle _{a}\right) \left| \overline{f}\right\rangle
_{v}\left| \overline{f}\right\rangle _{p}. 
\]

We can disregard the factor $\left| \overline{f}\right\rangle _{p}$, and
focus on the {\em quantum} part of the measurement outcome, $\left| \varphi
,t_{3}\right\rangle _{av}$, resulting from the reinterpretational step. We
should note that this reinterpretation, as it is, does not involve the
notion of time and is transparent to dynamics. Interestingly, the speed-up
stems out of the reinterpretation, i.e. by the constraint that there is only
one measurement outcome.

In fact, we will show that $\left| \varphi ,t_{3}\right\rangle _{av}$ is the
solution of a system of algebraic equations equivalent to (3). These
equations represent the following usual conditions introduced by quantum
measurement: \noindent (i) the outcome of measuring $\left[ v\right] $ must
be a single eigenstate $\left| f\right\rangle _{v}$, anyone element of the
set of all eigenstates $\left\{ \left| f\right\rangle _{v}\right\} $; (ii)
this eigenstate must ``drag'' all the tensor products appearing in $\left|
\varphi ,t_{2}\right\rangle _{av}$ that contain it; (iii) it must be a
specific eigenstate $\left| \overline{f}\right\rangle $, selected according
to probability amplitudes.

Let $\left| \varphi \right\rangle _{av}=%
\mathop{\textstyle\sum}%
_{x,y}\alpha _{x,y}\left| x\right\rangle _{a}\left| y\right\rangle _{v}$ be
an ``unknown'' vector of ${\cal H}_{av}$; $\left( x,y\right) $ runs over $%
B^{n}\times B^{n}$, and $\alpha _{x,y}$ are complex {\em variables}
independent of each other up to normalization: $%
\mathop{\textstyle\sum}%
_{x,y}\left| \alpha _{x,y}\right| ^{2}=1.$ The above conditions originate a
system of three algebraic equations to be simultaneously satisfied by $%
\left| \varphi \right\rangle _{av}$:

\begin{equation}
P_{v}^{f}\left| \varphi \right\rangle _{av}=\left| \varphi \right\rangle
_{av},
\end{equation}

\noindent where \noindent $P_{v}^{f}=\left| f\right\rangle _{v}\left\langle
f\right| _{v}$ is the projector on the Hilbert subspace ${\cal H}%
_{av}^{f}=span\left\{ \left| x\right\rangle _{a},\left| f\right\rangle
_{v}\right\} $ with $x$ running over $B^{n}$ and $\left| f\right\rangle
_{v}\in \left\{ \left| f\right\rangle _{v}\right\} $ being fixed; a $\left|
\varphi \right\rangle _{av}$ satisfying eq. (6) is a free linear combination
of all the tensor products of ${\cal H}_{av}$ containing $\left|
f\right\rangle _{v}$; this is condition (i);

\begin{equation}
\left| \left\langle \varphi \right. \right| _{av}\left| \left. \varphi
,t_{2}\right\rangle _{av}\right| \text{ must be maximum;}
\end{equation}

\noindent \noindent $\left| \varphi \right\rangle _{av}$, satisfying (6) and
(7) becomes the projection of $\left| \varphi ,t_{2}\right\rangle _{av}$ on $%
{\cal H}_{av}^{f}:\left| \varphi \right\rangle _{av}=\sqrt{\frac{N}{2}}%
\left| f\right\rangle _{v}\left\langle f\right| _{v}\left| \varphi
,t_{2}\right\rangle _{av};$ \noindent this means that $\left| f\right\rangle
_{v}$ has ``dragged'' all the tensor products of $\left| \varphi
,t_{2}\right\rangle _{av}$ containing it; this is condition (ii);

\begin{equation}
\left| f\right\rangle =\left| \overline{f}\right\rangle
\end{equation}

\noindent with $\left| \overline{f}\right\rangle $ randomly selected as
stated before.

The solution of equations (6-8) is $\left| \varphi \right\rangle _{av}=\sqrt{%
\frac{N}{2}}\left| \overline{f}\right\rangle _{v}\left\langle \overline{f}%
\right| _{v}\left| \varphi ,t_{2}\right\rangle _{av}=\left| \varphi
,t_{3}\right\rangle _{av}$, indeed the quantum state after measurement.

To sum up, satisfying equations (6-8) is equivalent to performing the
reinterpretational step of von Neumann's model. This is transparent to
measurement dynamics, namely to the first step of the model. Thus,
performing the first step gives ``for free'' (without incurring any further
dynamical cost) the solution of (6-8)\footnote{%
In any way, the process of satisfying equations (6-8) must be comprised in
the time interval $\left[ t_{2},t_{3}\right] $, which is linear in $n$. Ref. 
\cite{CAST99} provides a reformulation of von Neumann's model that better
fits the current approach.}. This is equivalent to solving equations (3),
namely the classically hard part of the problem. This justifies the quantum
speed-up.

The capability of directly solving a system of algebraic equations, without
having to execute an algorithm, comes from a peculiar feature. The
determination of the measurement outcome (i.e. of the solution) is {\em %
dually} {\em influenced} by both the initial actions, required to prepare
the state before measurement, and the logical-mathematical constraints
introduced by the final measurement action. These constraints are in fact 
{\em independent} of the initial actions since they hold unaltered for {\em %
all} initial actions.

In Simon's algorithm, dual influence is what distills a proper pair of
values of $x$ among an exponential number of such values, thus yielding the
speed-up. Conversely, the speed-up is the observable consequence of dual
influence.

Dual influence can be seen as a special instance of time-symmetrized quantum
measurement\footnote{%
The notion of time-symmetrized quantum measurement has been developed by
Aharanov et al., still outside the context of entanglement and problem
solving. See, e.g., refs. \cite{ABL64}, \cite{VAI98}.}. Whether this notion
is purely interpretational or can have observable consequences is a
controversial issue -- as well known. Unexpectedly, we have found a
certainly observable consequence (the speed-up) in the context of quantum
computation. It is worth noting that this consequence becomes observable 
{\em after} the action of quantum measurement and dual influence.

Summarizing, quantum computation turns out to belong to an entirely new
paradigm where there is identity between implicit or algebraic definition of
a solution and its physical determination.

It is worth noting that this paradigm blurs a long-standing distinction (of
mathematical logic) between the notions of ``implicit definition'' and
``computation''.

An implicit definition does not prescribe how to construct its object (say,
a string in some formal language). It only says that, demonstratedly, there
exists such an object. For example, the numerical problems we are dealing
with, implicitly or algebraically define their solutions\footnote{%
If the problem admits no solution, we should consider the meta-problem
whether the problem admits a solution.}. Let us consider factorization:
given the known product $c$ of two unknown prime numbers $x$ and $y$, the
numerical algebraic equation $x\cdot y=c$ implicitly defines the values of $%
x $ and $y$ that satisfy it. Equations (3) constitute a similar example.

In order to find the object of an implicit definition, the latter must be
changed into an equivalent constructive definition, namely into an algorithm
(if possible, but it is always possible with the problems we are dealing
with). An algorithm is an abstraction of the way things can be constructed
in reality -- inevitably in a model thereof -- and prescribes a computation
process that builds the object of the definition.

The current notion of algorithm still reflects the way things can be
constructed in the traditional classical reality -- namely through a
sequential process. Turing machine computation and the Boolean network
representation of computation are examples of sequential computation. An
algorithm specifies a one-way propagation of logical implication from a
completely defined input to a completely defined output which contains the
solution. It is thus meant to be executable through a dynamical process,
namely through a one-way causality propagation\footnote{%
Classical analog computation is not considered here to be fundamentally
different, being still performed through a one-way causality propagation.}.

We can see that the essence of quantum computation, dual influence, is
extraneous to the sequential notions of both algorithm and dynamics. In
particular, quantum computation {\em is not} ``quantum Turing machine''
computation.

\section{Four types of quantum algorithms}

\subsection{\noindent Modified Simon's algorithm}

In order to make our interpretation of the quantum speed-up more visible, we
will follow the simplified version\cite{EKAL97} of Simon's algorithm\cite
{SIM}. With respect to the original version, we must confine ourselves to
the case that the oracle gives us a 2-to-1 function $f:B^{n}\rightarrow
B^{n} $ such that

\[
\forall x\neq x^{^{\prime }}:f\left( x\right) =f(x^{^{\prime
}})\Longleftrightarrow x=x^{^{\prime }}\oplus r, 
\]
where $\oplus $ denotes bitwise exclusive or. The problem is to find $r$ in
poly(n) time. With a\ further simplification, as anticipated in Section II,
we replace the above condition with the condition $\left| x-x^{^{\prime
}}\right| =r$.

For the sake of clarity, the following table gives a trivial example.

\begin{center}
\begin{tabular}{|c|c|c|c|c|}
\hline
$x$ & $0$ & $1$ & $2$ & $3$ \\ \hline
$f\left( x\right) $ & $0$ & $1$ & $0$ & $1$ \\ \hline
\end{tabular}

Table I

\medskip
\end{center}

\noindent The modified algorithm is given in Fig. 2 -- we should disregard $%
/F$ for the time being.

\begin{center}
Fig. 2
\end{center}

\noindent Registers $a$ and $v$ undergo successive unitary transformations,
either jointly or separately:

\begin{itemize}
\item  The $f\left( x\right) $ transform (a reversible Boolean gate in the
time-diagram of computation -- Fig. 3) leaves the content of register $a$
unaltered, so that an input $x$ is repeated in the corresponding output, and
computes $f\left( x\right) $ adding it to the former content of register $v$
(which was set to zero). If the state is not sharp but is a quantum
superposition, the same transformation applies to any tensor product
appearing in it.
\end{itemize}

\begin{center}
\medskip Fig. 3
\end{center}

\begin{itemize}
\item  $H$ is the Hadamard transform. On a single qubit $i$, it operates as
follows: $\left| 0\right\rangle _{i}\stackrel{H}{\longrightarrow }\frac{1}{%
\sqrt{2}}\left( \left| 0\right\rangle _{i}+\left| 1\right\rangle _{i}\right)
,$ $\left| 1\right\rangle _{i}\stackrel{H}{\longrightarrow }\frac{1}{\sqrt{2}%
}\left( \left| 0\right\rangle _{i}-\left| 1\right\rangle _{i}\right) $. In
the general case of a register of n qubits, containing the number $\overline{%
x}$, it yields $\left| \overline{x}\right\rangle _{a}\stackrel{H}{\text{ }%
\longrightarrow \text{ }}\frac{1}{\sqrt{N}}\sum_{x}\left( -1\right) ^{%
\overline{x}\cdot x}\left| x\right\rangle _{a},$ where $N=2^{n}$, $x$ ranges
over $0,1,...$, $N-1$, and $\overline{x}\cdot x$ denotes the module 2 inner
product of the two numbers in binary notation (they should be seen as row
matrices).

\item  $M$ represents the action of measuring the numerical content of a
register.
\end{itemize}

The algorithm proceeds through the following steps (also applied to table I
example):

\medskip

\begin{enumerate}
\item[a)]  prepare:

$\left| \varphi ,t_{0}\right\rangle _{av}=\left| 0\right\rangle _{a}\left|
0\right\rangle _{v};$

perform the Hadamard transform on register $a$, this yields:

$\left| \varphi ,t_{1}\right\rangle _{av}=\frac{1}{\sqrt{N}}\sum_{x}\left|
x\right\rangle _{a}\left| 0\right\rangle _{v}=\frac{1}{2}\left( \left|
0\right\rangle _{a}\left| 0\right\rangle _{v}+\left| 1\right\rangle
_{a}\left| 0\right\rangle _{v}+\left| 2\right\rangle _{a}\left|
0\right\rangle _{v}+\left| 3\right\rangle _{a}\left| 0\right\rangle
_{v}\right) ;$

\item[c)]  compute $f\left( x\right) $ and add the result to the former
content $\left( 0\right) $ of register $v$, which yields:

$\left| \varphi ,t_{2}\right\rangle _{av}=\frac{1}{\sqrt{N}}\sum_{x}\left|
x\right\rangle _{a}\left| f\left( x\right) \right\rangle _{v}=\frac{1}{2}%
\left( \left| 0\right\rangle _{a}\left| 0\right\rangle _{v}+\left|
1\right\rangle _{a}\left| 1\right\rangle _{v}+\left| 2\right\rangle
_{a}\left| 0\right\rangle _{v}+\left| 3\right\rangle _{a}\left|
1\right\rangle _{v}\right) ;$ this is the state before measurement;

\item[d)]  measure $\left[ v\right] $ obtaining, say, $\overline{f}=1$; the
state after measurement is thus:

$\left| \varphi ,t_{3}\right\rangle _{av}=\frac{1}{\sqrt{2}}\left( \left| 
\overline{x}\right\rangle _{a}+\left| \overline{x}+r\right\rangle
_{a}\right) \left| \overline{f}\right\rangle _{v}=\frac{1}{\sqrt{2}}\left(
\left| 1\right\rangle _{a}+\left| 3\right\rangle _{a}\right) \left|
1\right\rangle _{v},$

We should note that, at this stage of the algorithm, it is {\em equivalent}
to either perform or skip $\left[ v\right] $ measurement (see further
below). It will be easier to understand the algorithm and the reason of the
speed-up if we assume that this measurement has been performed. The
measurement outcome, $\left| \varphi ,t_{3}\right\rangle _{av}=\sqrt{\frac{N%
}{2}}\left| \overline{f}\right\rangle _{v}\left\langle \overline{f}\right|
_{v}\left| \varphi ,t_{2}\right\rangle _{av}$, is naturally \ dually
influenced (Section III).
\end{enumerate}

\medskip

Ekert and Jozsa\cite{EKJ98} have shown that quantum entanglement between
qubits is essential for providing a computational speed up, in terms of time 
{\em or} resources, in the class of quantum algorithms we are dealing with
(which yield an exponential speed up). After measuring $f\left( x\right) $,
the state of the two registers becomes factorizable, and all entanglement is
destroyed. The remaining actions, performed on register $a$, use
interference (which generates no entanglement) to ``extract'' $r$ out of the
superposition $\frac{1}{\sqrt{2}}\left( \left| \overline{x}\right\rangle
_{a}+\left| \overline{x}+r\right\rangle _{a}\right) $. Under the criterion
introduced in Section II, we must conclude from another standpoint that the
speed-up has been achieved by preparing $\left| \varphi ,t_{3}\right\rangle
_{av}$.

\medskip

\begin{enumerate}
\item[e)]  perform $H$ on register $a$, this yields:

$\left| \varphi ,t_{4}\right\rangle _{av}=\frac{1}{\sqrt{2N}}\sum_{z}\left(
-1\right) ^{\overline{x}\cdot z}\left[ 1+\left( -1\right) ^{r\cdot z}\right]
\left| z\right\rangle _{a}\left| \overline{f}\right\rangle _{v}$;

\item[f)]  measure $\left[ a\right] $ in $\left| \varphi ,t_{4}\right\rangle
_{av}$; we designate the result by $z$;

$r\cdot z$ must be 0 -- see the form of $\left| \varphi ,t_{4}\right\rangle
_{av}$. This holds unaltered if step (d) measurement is omitted, as well
known;

\item[g)]  by repeating the overall computation process a sufficient number
of times, poly($n$) on average, a number of constraints $r\cdot z=0$
sufficient to identify $r$ is gathered.
\end{enumerate}

\medskip

How the speed-up is achieved in $\left[ t_{0},t_{3}\right] $ has been
anticipated in Sections II and III. Summarizing, measuring $\left[ v\right] $
in state $\left| \varphi ,t_{2}\right\rangle _{av}$, creates the system of
algebraic equations (6-8) [equivalent to (3)] and yields the superposition
of a pair of values of $x_{1}$ and $x_{2}$ which satisfy this system ($r$ is
``easily'' extracted from the superposition). Solving equations (3) by
classical computation would require exp($n$) time.

Finally, let us show that performing or skipping step (d) (i.e. $\left[ v%
\right] $ measurement in $\left| \varphi ,t_{2}\right\rangle _{av}$) is {\em %
equivalent}. Let us skip step (d) and measure $\left[ a\right] $ first, at
time $t_{4}$. In Fig. 2, $M$ on $v$ should be shifted at least after $t_{5}$%
. Whether $\left[ v\right] $ is measured after $t_{5}$ is indifferent, or
mathematically equivalent. Let us think of measuring it. This induces a
``wave function collapse''\footnote{%
The notion of ``collapse'' is not needed in any essential way; it is a
mathematically legitimate notion that comes handy here for the sake of
explanation; the result of collapse can be backdated any time during the
unobserved evolution of the quantum system from $t_{0}$ to $t_{3}$, provided
that this result undergoes back in time (in an inverted way) the same
transformations undergone by the time-forward evolution (the usual one).} of
the state of register $v$ on some $\left| \overline{f}\right\rangle _{v}$.
Since $\left| \overline{f}\right\rangle _{v}$ is disentangled from the state
of register $a$, and no operation has been performed on register $v$ since
time $t_{2}$ (see fig. 2, keeping in mind that $M$ on $v$ has been shifted
after $t_{5}$), back-dating collapse at time $t_{2}$ means back-dating the
result of collapse, namely $\left| \overline{f}\right\rangle _{v}$, as it
is. This is equivalent to having performed step (d). Another way of seeing
this is that, because of the entanglement between registers $a$ and $v$,
measuring $\left[ a\right] $ first, at time $t_{4}$, is equivalent to
simultaneously measuring $\left[ v\right] $; the result of this virtual
measurement can be backdated, and we can go on with a reasoning similar to
the above one.

\subsection{Shor's algorithm}

\noindent The problem of factoring an integer $L$ -- the product of two
unknown primes -- is transformed into the problem of finding the period of
the function $f\left( x\right) =a^{x}%
\mathop{\rm mod}%
L$, where $a$ is an integer between $0$ and $L-1$, and is coprime with $L$%
\cite{SHOR}, \cite{EKAL99}. Figure 2 can also represent Shor's algorithm,
provided that $f\left( x\right) $ is defined as above and that the second
Hadamard transform is substituted by the discrete Fourier transform $F$. The
state before measurement has the form $\left| \varphi ,t_{2}\right\rangle
_{av}=\frac{1}{\sqrt{L}}\sum_{x}\left| x\right\rangle _{a}\left| f\left(
x\right) \right\rangle _{v}$. Measuring or not measuring $f\left( x\right) $
in $\left| \varphi ,t_{2}\right\rangle _{av}$ is still equivalent. By
measuring it, the above quantum state changes into the superposition

\begin{equation}
\overline{k}\left( \left| \overline{x}\right\rangle _{a}+\left| \overline{x}%
+r\right\rangle _{a}+\left| \overline{x}+2r\right\rangle _{a}+...\right)
\left| \overline{f}\right\rangle _{v},
\end{equation}

\noindent where $f\left( \overline{x}\right) =f\left( \overline{x}+r\right)
=...=\overline{f}$, and $\overline{k}$ is a normalization factor.

The second part of the algorithm generates no entanglement and serves to
``extract'' $r$ in polynomial time, by using Fourier-transform interference
and auxiliary, off line, mathematical considerations. Under the current
assumptions, the quantum speed-up has been achieved by preparing state (9):
the discussion is completely similar to that of the previous algorithm.

\subsection{Deutsch's 1985 algorithm}

\noindent The seminal 1985 Deutsch's algorithm has been the first
demonstration of a quantum speed-up. In its current form, this algorithm
yields a deterministic output, apparently ruling out the dual influence
explanation. A\ thorough examination of both the problem and the solution
algorithm will show that this is not the case.

Until now, the problem has been to efficiently reverse a hard-to-reverse
function $f\left( x\right) $. In the language of game theory, this is a game
against (mathematical) nature. Deutsch's algorithm and more in general
quantum oracle computing is better seen as a competition between two
players. One produces the problem, the other should produce the solution.
Sticking to Greek tradition, we shall call the former player Sphinx, the
latter Oedipus.

The game is formalized as follows. Both players know everything of a set of
software programs $\left\{ f_{k}\right\} $ (where $k$ labels the elements of
the set), whereas each program $f_{k}$ computes some function $%
f_{k}:B^{n}\rightarrow B^{n}$. The Sphinx chooses $k$ at random, loads
program $f_{k}$ on a computer (i.e., sets the oracle in its $k$-th mode) and
passes it on to Oedipus. Oedipus knows nothing of the Sphinx' choice and
must efficiently find $k$ by testing the computer (oracle) input-output
behaviour. If the computer is quantum, then we speak of ``quantum oracle
computing''.

Deutsch's 1985 algorithm\cite{DEUTSCH}, as modified in \cite{EKAL97}, is as
follows. Let \noindent $\left\{ f_{k}\right\} $ be the set of all possible
functions \noindent $f_{k}:B\rightarrow B$, namely:

\medskip

\begin{center}
\begin{tabular}{|c|c|cc|c|c|cc|c|c|cc|c|c|}
\cline{1-2}\cline{5-6}\cline{9-10}\cline{13-14}
$x$ & $f_{00}\left( x\right) $ & \qquad &  & $x$ & $f_{01}\left( x\right) $
& \qquad &  & $x$ & $f_{10}\left( x\right) $ & \qquad &  & $x$ & $%
f_{11}\left( x\right) $ \\ \cline{1-2}\cline{5-6}\cline{9-10}\cline{13-14}
$0$ & $0$ & \qquad &  & $0$ & $0$ & \qquad &  & $0$ & $1$ & \qquad &  & $0 $
& $1$ \\ \cline{1-2}\cline{5-6}\cline{9-10}\cline{13-14}
$1$ & $0$ & \qquad &  & $1$ & $1$ & \qquad &  & $1$ & $0$ & \qquad &  & $1 $
& $1$ \\ \cline{1-2}\cline{5-6}\cline{9-10}\cline{13-14}
\end{tabular}
\end{center}

\noindent \medskip

\noindent $\left\{ f_{k}\right\} $ is divided into a couple of subsets: the
balanced functions, characterized by an even number of zero and one values,
thus labeled by $k=01,10$, and the unbalanced ones, labeled by $k=00,11$.
Once set in its $k$-th mode, the oracle computes $f_{k}\left( x\right) $.
Oedipus must find, with a minimum number of oracle runs, whether the oracle
(whose mode has been randomly set by the Sphinx) computes a balanced or an
unbalanced function. In other words, he must compute the functional ${\cal F}%
\left( f_{k}\right) $ which is, say, 1 (0) when the function is balanced
(unbalanced). The algorithm is illustrated in Fig. 4(a). The computation of $%
f_{k}\left( x\right) $ is represented as a reversible Boolean gate like in
the previous algorithms, but for the fact that the result of the computation
is now module 2 added to the former content of register $v$.

\begin{center}
Fig. 4(a),(b)
\end{center}

Given the Sphinx' choice $k$, the algorithm proceeds as follows; each point
gives the action and the corresponding result.

\begin{enumerate}
\item[a)]  prepare:

$\left| \varphi _{k},t_{0}\right\rangle _{av}=\frac{1}{\sqrt{2}}\left|
0\right\rangle _{a}\left( \left| 0\right\rangle _{v}-\left| 1\right\rangle
_{v}\right) ,$

\item[b)]  perform Hadamard on $a$:

$\left| \varphi _{k},t_{1}\right\rangle _{av}=\frac{1}{2}\left( \left|
0\right\rangle _{a}+\left| 1\right\rangle _{a}\right) \left( \left|
0\right\rangle _{v}-\left| 1\right\rangle _{v}\right) ,$

\item[c)]  we shall consolidate the next two steps -- Fig. 4(a): compute $%
f_{k}\left( x\right) $ adding it, module 2, to the former content of $v$,
and perform Hadamard on $a$; the result depends on the Sphinx' choice:

$\left| \varphi _{00},t_{3}\right\rangle _{av}=\frac{1}{\sqrt{2}}\left|
0\right\rangle _{a}$ $\left( \left| 0\right\rangle _{v}-\left|
1\right\rangle _{v}\right) $

$\left| \varphi _{01},t_{3}\right\rangle _{av}=\frac{1}{\sqrt{2}}\left|
1\right\rangle _{a}$ $\left( \left| 0\right\rangle _{v}-\left|
1\right\rangle _{v}\right) $

$\left| \varphi _{10},t_{3}\right\rangle _{av}=-\frac{1}{\sqrt{2}}\left|
1\right\rangle _{a}$ $\left( \left| 0\right\rangle _{v}-\left|
1\right\rangle _{v}\right) $

$\left| \varphi _{11},t_{3}\right\rangle _{av}=-\frac{1}{\sqrt{2}}\left|
0\right\rangle _{a}$ $\left( \left| 0\right\rangle _{v}-\left|
1\right\rangle _{v}\right) $

\item[d)]  measure $\left[ a\right] $: it can be seen that the content of
register $a$ yields the functional ${\cal F}\left( f_{k}\right) $, namely
Oedipus' answer.
\end{enumerate}

This algorithm is more efficient than any classical algorithm, where two
runs of the oracle are required to compute ${\cal F}\left( f_{k}\right) $.
However, the result is apparently reached in a deterministic way, without
any\ active role of quantum measurement.

This must be ascribed to an incomplete physical representation of the
problem. In Section III, we had a problem that implicitly defined its
solution, whereas this mathematical fact was physically represented by the
quantum measurement of an entangled state. This obviously requires that the
problem is physically represented\footnote{%
In Sections IV.A and IV.B, all knowledge of the function and ignorance about 
$r$ were physically represented in a superposition of the form (1).},
whereas presently an essential part of it, the Sphinx choosing the oracle
mode, is not.

First, we shall follow a most simple way of completing the physical
representation. The Sphinx' random selection of the oracle mode will be
performed through a suitable quantum measurement, after having run the
algorithm.

We introduce the extended gate $F\left( k,x\right) $ which computes the
function $F\left( k,x\right) =f_{k}\left( x\right) $ for all $k$ and $x$.
This gate has an ancillary input register $m$ ($m$ for mode) which contains $%
k$, namely the oracle mode [Figure 4(b) gives the extended algorithm]. This
input is identically repeated in a corresponding output -- to keep gate
reversibility. Of course, Oedipus is forbidden to access register $m$. The
preparation becomes 
\[
\left| \varphi ,t_{0}\right\rangle _{mav}=\frac{1}{\sqrt{2}}\left|
00\right\rangle _{m}\left| 0\right\rangle _{a}\left( \left| 0\right\rangle
_{v}-\left| 1\right\rangle _{v}\right) . 
\]
After performing Hadamard on $m$ and $a$ we obtain:

\begin{equation}
\left| \varphi ,t_{1}\right\rangle _{mav}=\frac{1}{4}\left( \left|
00\right\rangle _{m}+\left| 01\right\rangle _{m}+\left| 10\right\rangle
_{m}+\left| 11\right\rangle _{m}\right) \left( \left| 0\right\rangle
_{a}+\left| 1\right\rangle _{a}\right) \left( \left| 0\right\rangle
_{v}-\left| 1\right\rangle _{v}\right) .
\end{equation}

\noindent Performing Hadamard on $\left| 00\right\rangle _{m}$ is a way of
preparing the Sphinx' random selection of an oracle mode (as will become
clear). Let us go directly to the state before the first measurement -- see
Fig. 4(b)

\begin{equation}
\left| \varphi ,t_{3}\right\rangle _{mav}=\frac{1}{2\sqrt{2}}\left[ \left(
\left| 00\right\rangle _{m}-\left| 11\right\rangle _{m}\right) \left|
0\right\rangle _{a}+\left( \left| 01\right\rangle _{m}-\left|
10\right\rangle _{m}\right) \left| 1\right\rangle _{a}\right] \left( \left|
0\right\rangle _{v}-\left| 1\right\rangle _{v}\right) .
\end{equation}

It can be seen that the entangled state (11) represents the {\em mutual
definition} between the Sphinx' choice $k$ and Oedipus' answer ${\cal F}%
\left( f_{k}\right) $. The former implicit definition of the problem
solution appears here in the form of the mutual definition of the moves of
the two players. Reaching state (11) with quantum parallel computation still
requires one oracle run.

The action of measuring $\left[ m\right] $ in state (11), equivalent to the
Sphinx' choice of the oracle mode, by bringing in and satisfying equations
(6-8)\footnote{%
The ``state before measurement'' $\left| \varphi ,t_{2}\right\rangle _{av}$
of Section III must be changed into $\left| \varphi ,t_{3}\right\rangle
_{mav}$.}, transforms mutual definition into correlation between individual
outputs (like in an EPR situation). In other words, the Sphinx' choice of $k$
simultaneously determines Oedipus' answer ${\cal F}\left( f_{k}\right) $ --
retrievable by measuring $\left[ a\right] $. In the classical framework
instead, the Sphinx' choice should necessarily be propagated to Oedipus'
answer by means of an algorithm, in fact through the computation of ${\cal F}%
\left( f_{k}\right) $ (requiring two oracle runs).

Achieving the speed-up still involves the interplay between the reversible
preparation of an entangled state before measurement and a final measurement
action, namely dual influence -- here of an EPR kind.

It should be noted that the above ``complete physical representation'' is
not the original Deutsch's algorithm. This can readily be fixed. To this
end, the Sphinx must randomly select the mode before giving the oracle --
i.e. the quantum gate $F\left( k,x\right) $ -- to Oedipus. This means that
Oedipus receives the oracle in an input state randomly selected among four
possible quantum states, corresponding to the modes $k=00,01,10,11.$ This is
indistinguishable from a mixture. Therefore, the preparation at time $t_{1}$
becomes:

\[
\left| \varphi ,t_{1}\right\rangle _{mav}=\frac{1}{4}\left( \left|
00\right\rangle _{m}+e^{i\delta _{1}}\left| 01\right\rangle _{m}+e^{i\delta
_{2}}\left| 10\right\rangle _{m}+e^{i\delta _{3}}\left| 11\right\rangle
_{m}\right) \left( \left| 0\right\rangle _{a}+\left| 1\right\rangle
_{a}\right) \left( \left| 0\right\rangle _{v}-\left| 1\right\rangle
_{v}\right) , 
\]

\noindent where $\delta _{1}$, $\delta _{2}$ and $\delta _{3}$ are
independent random phases -- this is the random phase representation of a
mixture\cite{FINK}. After $t_{1}$, the algorithm goes on as before yielding

\[
\left| \varphi ,t_{3}\right\rangle _{mav}=\frac{1}{2\sqrt{2}}\left[ \left(
\left| 00\right\rangle _{m}-e^{i\delta _{3}}\left| 11\right\rangle
_{m}\right) \left| 0\right\rangle _{a}+\left( e^{i\delta _{1}}\left|
01\right\rangle _{m}-e^{i\delta _{2}}\left| 10\right\rangle _{m}\right)
\left| 1\right\rangle _{a}\right] \left( \left| 0\right\rangle _{v}-\left|
1\right\rangle _{v}\right) . 
\]

\noindent Clearly, the roles of entanglement, quantum measurement and dual
influence remain unaltered.

\subsection{An instance of Grover's algorithm}

\noindent The rules of the game are the same as before. This time we have
the set of the $2^{n}$ functions $f_{k}:B^{n}\rightarrow B$ such that $%
f_{k}\left( x\right) =\delta _{k,x}$, where $\delta $ is the Kronecker
symbol. We shall consider the simplest instance $n=2$. This yields four
functions $f_{k}\left( x\right) $, labeled by $k=0,1,2,3$. Figure 5(a) gives
Grover's algorithm\cite{GROVER} (in the standard version provided in \cite
{EKAL97} for $n=2$. Let us assume the Sphinx has chosen $k=2$. The
preparation is $\frac{1}{\sqrt{2}}\left| 0\right\rangle _{a}\left( \left|
0\right\rangle _{v}-\left| 1\right\rangle _{v}\right) $. Without entering
into detail, the state before measurement is: $\frac{1}{\sqrt{2}}\left|
2\right\rangle _{a}\left( \left| 0\right\rangle _{v}-\left| 1\right\rangle
_{v}\right) $. Measuring $\left[ a\right] $ deterministically yields
Oedipus' answer. This is more efficient than classical computation where
three oracle runs are required to find the solution with certainty, whereas
in Grover's algorithm two runs are enough -- Fig. 5(a).

\begin{center}
Fig. 5(a),(b)
\end{center}

The extended algorithm is given in Fig. 5(b). The preparation becomes

$\frac{1}{\sqrt{2}}\left| 0\right\rangle _{m}\left| 0\right\rangle
_{a}\left( \left| 0\right\rangle _{v}-\left| 1\right\rangle _{v}\right) $;

the state before measurement becomes:

$\frac{1}{2\sqrt{2}}\left( \left| 0\right\rangle _{m}\left| 0\right\rangle
_{a}+\left| 1\right\rangle _{m}\left| 1\right\rangle _{a}+\left|
2\right\rangle _{m}\left| 2\right\rangle _{a}+\left| 3\right\rangle
_{m}\left| 3\right\rangle _{a}\right) \left( \left| 0\right\rangle
_{v}-\left| 1\right\rangle _{v}\right) $.

Again, we have the mutual definition of the Sphinx' choice and Oedipus'
answer. Measuring $\left[ m\right] $ selects the Sphinx' choice and Oedipus'
answer at the same time, as in the previous oracle problem.

\section{Conclusions}

\noindent Quantum computation is concerned with the efficient solution of
numerical algebraic problems. We shall first summarize the main results of
this work.

We have shown that the action of measuring an observable in a suitably
entangled state, introduces and satisfies a system of algebraic equations.
In all existing quantum algorithms, this system represents the problem that
algebraically defines its solution. Moreover, measurement time is
independent of entanglement. This justifies the quantum speed-up in all
types of quantum algorithms found so far. Quantum computation turns out to
be an entirely new paradigm (extraneous to the notion of sequential
computation) where there is identity between the algebraic definition of a
solution and its physical determination.

The capability of directly solving\footnote{%
Without having to execute an algorithm, which would be necessary in the
classical framework.} a system of algebraic equations, is related to the
feature that the determination of the measurement outcome is dually
influenced by both the reversible initial actions, leading to the state
before measurement, and by the logical-mathematical constraints introduced
by the final measurement action. Dual influence is extraneous to the notion
of sequential process, namely of dynamical, one-way propagation.

Although our explanation of the speed-up appears a-posteriori to be simple
and evident, it is likely to displace rather common views. In the first
place, it is reasonable to assume that quantum algorithms are commonly
thought to be, in fact, algorithms, namely the quantum transposition of
sequential Turing machine. At the light of the results of this work, this
way of thinking would be a classical vestige, ruling out the active role of
quantum measurement and dual influence.

In the second place, there is a widespread belief that quantum theory can do
without the measurement problem. In other words, the fact that the mutual
exclusivity of the possible measurement outcomes comes from an ad-hoc
reinterpretation of a state superposition (of a mixture, in decoherence
theory) would be a price paid once for all. There would be no further
consequences on quantum theory. In contrast with this, we have highlighted a
striking consequence in the context of quantum computation. Here the
``reinterpretation'' implies dual influence, which yields a completely
observable speed-up.

These appear to be important clarifications provided by this work.

From the one hand, the notion of dual influence, with its striking
consequence, might lend itself to further development at a fundamental level.

From the other hand, having ascertained that quantum algorithms are more
than sequential computation, might open the way to unforeseen prospects in
the quest of new forms of computation. For example, quantum measurement of
an observable in an entangled state is a {\em projection }on a Hilbert
subspace subject to certain constraints whose satisfaction amounts to
efficiently solving a problem. In some respect, this feature is similar to
the projections due to particle statistics symmetrizations. Therefore,
investigating the possibility of exploiting such symmetrizations in problem
solving could be an interesting prospect. Refs. \cite{PHYSD}, \cite{LIPT}
provide still abstract attempts in this direction.

More generally, this work highlights the essential role played by
non-dynamical effects in quantum computation. Let us mention in passing that
a form of quantum computation which is of geometric rather than dynamical
origin has recently been provided \cite{JONES}. This concretely shows that
there are ways of getting out of the usual quantum computation paradigm.

\smallskip

Thanks are due to T. Beth, A. Ekert, D. Finkelstein and V. Vedral for
stimulating discussions and valuable comments.

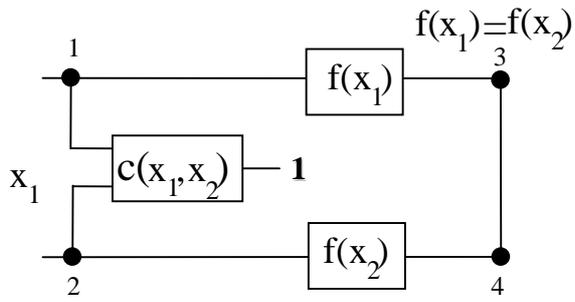
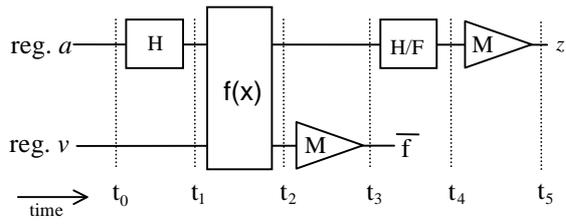
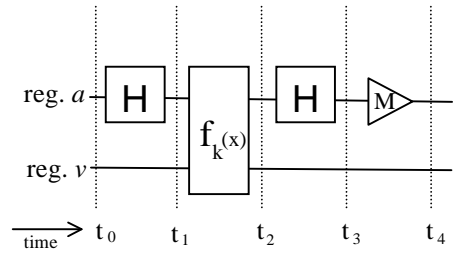
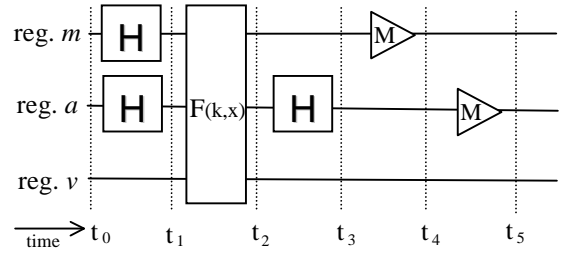
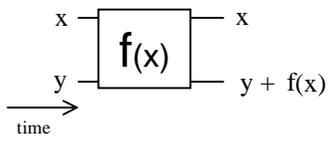
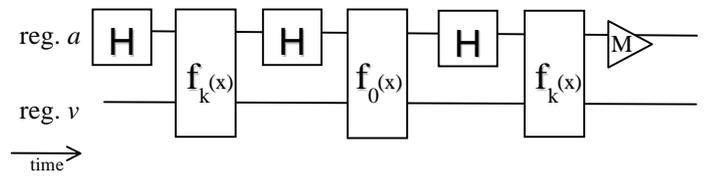
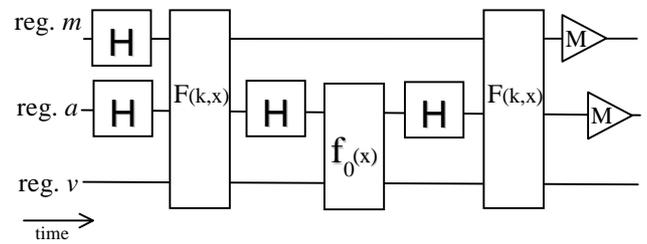
\end{document}